\documentclass[10pt,twocolumn,superscriptaddress,english,10pt,prl,showpacs,floatfix,aps]{revtex4-1}
\usepackage[utf8]{inputenc}
\usepackage{amsthm}
\usepackage{amsmath}
\usepackage{amssymb}
\usepackage{esint}
\usepackage{graphicx}
\usepackage{upgreek}
\usepackage{gensymb}

\newcommand{\Figref}[1]{Figure~\ref{#1}}

\begin{document}

\title{Disentangling electron- and electric field-induced ring-closing reactions in a diarylethene derivative on Ag(111)}

\author{Ga\"el Reecht}\email{greecht@zedat.fu-berlin.de}
\affiliation{ Fachbereich Physik, Freie Universit\"{a}t Berlin, Arnimallee 14, 14195 Berlin, Germany}
\author{Christian Lotze}
\affiliation{ Fachbereich Physik, Freie Universit\"{a}t Berlin, Arnimallee 14, 14195 Berlin, Germany}
\author{Dmytro Sysoiev}
\affiliation{ Fachbereich Chemie, Universit\"{a}t Konstanz, Universit\"atsstra\ss e 10, 78457 Konstanz, Germany}
\author{Thomas Huhn}
\affiliation{ Fachbereich Chemie, Universit\"{a}t Konstanz, Universit\"atsstra\ss e 10, 78457 Konstanz, Germany}
\author{Katharina J Franke}
\affiliation{ Fachbereich Physik, Freie Universit\"{a}t Berlin, Arnimallee 14, 14195 Berlin, Germany}

\begin{abstract}
Using scanning tunneling microscopy and spectroscopy we investigate the adsorption properties and ring-closing reaction of a diarylethene derivative (C5F-4Py) on a Ag(111) surface. We identify an electron-induced reaction mechanism, with a quantum yield varying from $10^{-14}-10^{-9}$ per electron upon variation of the bias voltage from $1-2$~V. We ascribe the drastic increase in switching efficiency to a resonant enhancement upon tunneling through molecular orbitals. Additionally, we resolve the ring-closing reaction even in the absence of a current passing through the molecule. In this case the electric-field can modify the reaction barrier, leading to a finite switching probability at 4.8~K. A detailed analysis of the switching events shows that a simple plate-capacitor model for the tip-surface junction is insufficient to explain the distance dependence of the switching voltage. Instead, describing the tip as a sphere is in agreement with the findings. We resolve small differences in the adsorption configuration of the closed isomer, when comparing the electron- and field-induced switching product.

\end{abstract}

\maketitle

\section{Introduction}

Diarylethene (DAE) molecules are prototypes of molecular ring-opening/closing switches \cite{Irie2000, Tian2004,Irie2014}. In the ring-closed form the molecules are fully-$\pi$-conjugated, whereas the conjugation is interrupted in the open form. Therefore, the two isomers exhibit drastic differences in their optical and conductance properties, while their geometries are only slightly changed. The optical absorption spectrum moves from the ultra-violet range in the open form to the visible range in the closed isomer \cite{Irie2014,Gilat1995}. The conductance ratio, which is an important figure of merit as the ON/OFF ratio, is in the order of 10 to 300 depending on the endgroups \cite{Dulic2003, Kim2012, Jia2013, Jia2016, Reecht2016}.
The intriguing combination of structural, optical and electronic properties renders these molecules highly-promising candidates for their application in electronic and opto-electronic devices \cite{Whalley2007, Song2011, Kim2014}. Device applications also require reliable means to switch between the two states. Incorporation of the molecules into device geometries, however, crucially modifies the properties of the individual isomers and often deteriorates the switching ability \cite{VanderMolen2010, Tegeder2012}. Fascinating results of reversible light-induced switching at room temperatures has recently been achieved in DAEs contacted to graphene electrodes \cite{Jia2016}.

Despite of the huge progress made in light-triggered switching, it is still necessary to consider other stimuli. This is especially important for  applications in single-molecule devices, because of the fundamental limits of light focusing. In contrast, charge flow through molecules can be limited to the individually addressable molecule. In fact, electron-induced switching is widely investigated for switching molecules on metal electrodes \cite{VanderMolen2010,Morgenstern2011, Kim2002, Choi2006, Liljeroth2007, Henningsen2007, Henzl2010, Lotze2012,Ladenthin2015}. Voltage biasing the molecules to control the current inevitably includes the presence of an electric field. The electric field by itself may affect the potential energy landscape of the molecular isomers and thus modify or induce/hinder a switching process \cite{Alemani2006, Wirth2015}. It is therefore important to understand both, the electric field as well as the electron-induced switching processes, and the interplay of the effects. 
				
\begin{figure}[ht]
\begin{center}
\includegraphics[width=1\columnwidth]{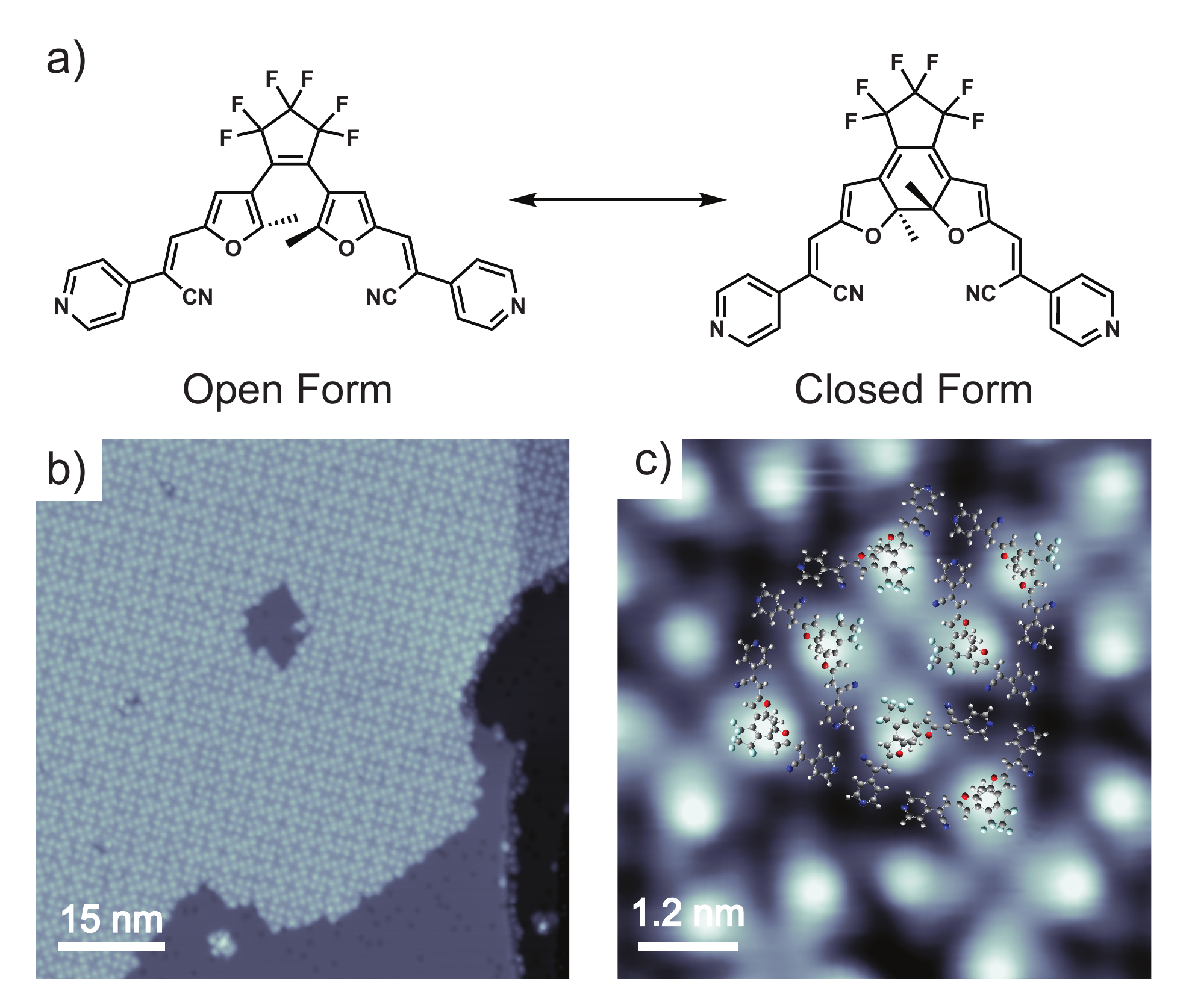}
\end{center}
\caption{ a) Chemical structure of the open (left) and closed (right) forms of the C5F-4Py molecule. b) Large-scale  STM image (75~$\times$~75~nm$^{2}$) of the Ag(111) surface with molecular islands of C5F-4Py (V~=~0.8~V, I~=~32~pA). c) High resolution STM image (8~$\times$~8~nm$^{2}$) of a molecular island (V~=~0.5~V, I~=~120~pA). The model of the C5F-4Py molecule calculated in gas phase is overlaid to indicate the proposed assembly structure.  }
\label{fig1}
\end{figure}

To elucidate the electron and electric-field effects on a ring-closing reaction, we investigated 1,2-bis(2-methyl-5-((Z)-(2-cyano-2-(pyridin-4-yl)vinyl))furan-3-yl) hexafluorocyclopentene (C5F-4Py) \cite{Sysoiev2011} (\Figref{fig1}a)) on a Ag(111) surface with  low temperature scanning tunneling microscopy (STM) and scanning tunneling spectroscopy (STS). This DAE-derivative is functionalized with oxygen containing furyl and therefore expected to interact less with a metal than the more common parent compound, which contains sulfur bearing thiophenyl aryl-rings \cite{Kim2012,Reecht2016}. Adsorbed on Ag(111), the molecules self-assemble in islands. We show that the open form can be switched into the closed form by the STM tip. We identify an electron-induced as well as an electric-field induced reaction. These processes can be disentangled by current- and height dependent determination of the switching efficiency. The quantum yield for electron-induced switching is resonantly enhanced when tunneling through the LUMO derived orbitals. Our detailed analysis further reveals that the typically assumed plate-capacitor model is insufficient to describe the distance dependence of the switching voltage. We show that the next higher approximation of a spherical tip is a better model to describe the behavior.

\section{Experimental details}

The C5F-4Py diarylethene was synthesized according to Ref.6.
The Ag(111) sample was cleaned by repeated Ne$^+$ sputtering and annealing cycles. The open C5F-4Py isomers were evaporated from a Knudsen cell at 440~K onto the Ag(111) surface held at room temperature in ultrahigh vacuum. The tungsten STM tip was covered with silver by indentation into the clean surface. The quality of the tip was checked on the clean surface by observing the silver surface state in spectroscopy.
All STM measurements were performed with a home-built STM at a temperature of 4.8~K. Differential conductance spectra were recorded with an open feedback-loop using lock-in detection with a modulation frequency of 910~Hz and a root mean square modulation amplitude of 15~mV.

\section{Results and discussion}

\subsection{Molecular adsorption structure}

\begin{figure}[ht]
\begin{center}
\includegraphics[width=1\columnwidth]{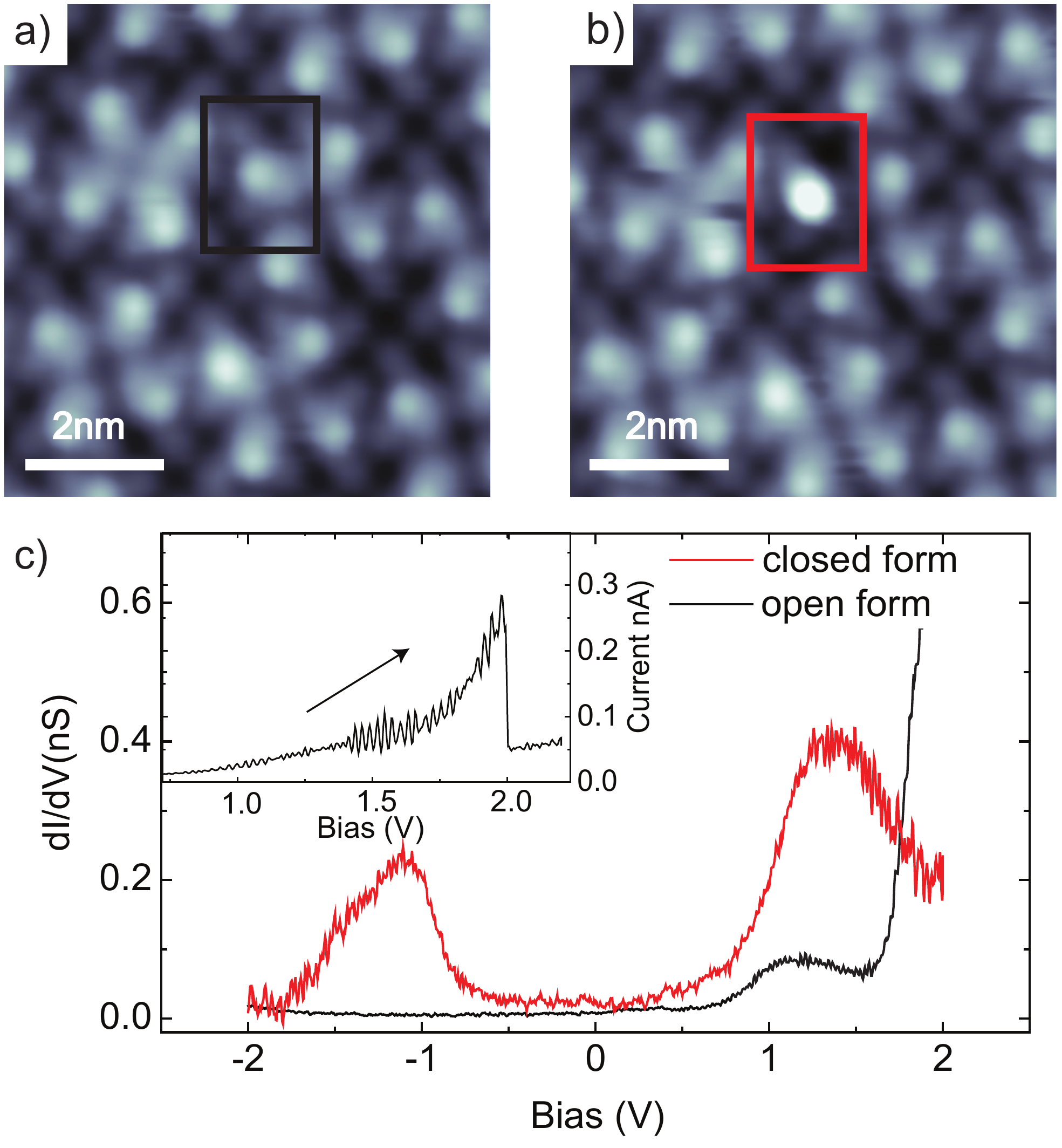}
\end{center}
\caption{ STM topographies before (a) and after (b) a voltage ramp has been applied on the top of the molecule in the rectangle (V~=~-1.3~V, I~=~50~pA). c) $dI/dV$ spectra recorded on the top of the framed molecules in a) and b), before (black) and after (red) the voltage ramp, corresponding respectively to the open and closed form of the C5F-4Py molecules. The trace of the current during the voltage ramp inducing the switching is shown as inset.}
\label{fig2}
\end{figure}

Evaporation of the open isomer of C5F-4Py from the powder onto a clean Ag(111) surface at room temperature leads to densely-packed molecular islands of a monolayer height (\Figref{fig1}b). High resolution STM images as in \Figref{fig1}c allow for an assignment of the individual molecules. Similar to their appearance on Au(111) \cite{Reecht2016}, each molecule is roughly of triangular shape with a brighter feature corresponding to the switching unit of the DAE, and two lower parts associated to the cyano-vinyl-pyridine groups of the molecule. We find four orientations of the molecules  rotated by  90\,\degree  with respect to each other.

An important observation is that with our evaporation conditions, all molecules appear similar in shape indicating that they are all in the same isomeric state. However, a modification of the appearance of the molecule can be induced deliberately by the STM tip as revealed by a comparison of the topographies in \Figref{fig2}a and \Figref{fig2}b, where one molecule appears brighter in the second image. This kind of change is obtained by applying a voltage ramp over the molecule until the simultaneously recorded current drops drastically (inset in \Figref{fig2}c). 

The differential conductance spectra ($dI/dV$) of the molecule before (black) and after (red) the modification are shown in \Figref{fig2}c. The initial molecule exhibits  two resonances at positive bias ($\sim$~1.2~V and $\sim$~2~V) and none at negative bias in the range down to -2\,V. The spectrum of the bright molecule is drastically different with one resonance at positive bias ($\sim$~1.3~V) and one at negative ($\sim$~-1.2~V). The latter resonance dominates the STM topography in \Figref{fig2}b and allows for the reliable and fast identification of the isomeric state, which will be important for the following experiments. 
Similar changes in the electronic structure were also observed for C5F-4Py molecules on Au(111) and associated to the switching from the open to the closed isomer \cite{Reecht2016}. The reduction of the HOMO-LUMO gap is in agreement with the change from a non-conjugated (open form) to a fully conjugated molecule (closed form). The presence of two unoccupied states with similar energies in the initial molecule can be explained by the lifting of the degeneracy of the LUMO state of the open isomer \cite{Reecht2016}. The switching described above was reproducible on hundreds of molecules. However, we could not observe a back-switching from the closed to the open form.

\subsection{Electron-induced switching}
\label{sec_elinduced}

\begin{figure*}[ht]
\begin{center}
\includegraphics[width=2\columnwidth]{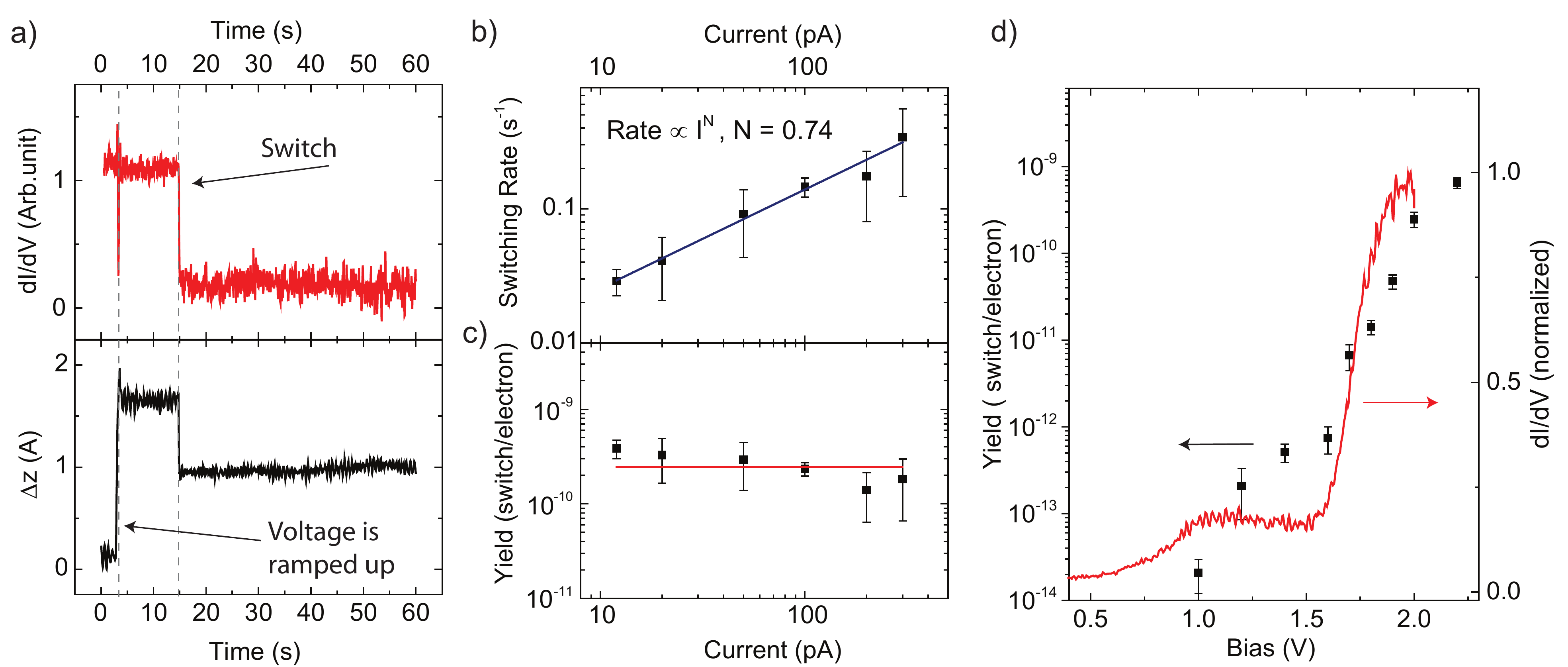}
\end{center}
\caption{ a) Constant-current (50\,pA) time traces of the $dI/dV$ signal (red) and the topography (black) on the center of an open molecule during a voltage pulse of 2~V, 60~s. b) Switching rate $R$ versus current at a sample bias of 2~V. The full line is a fit to the rate $R=a\times I^N$. c) Switching yield versus current at a sample bias of 2~V. The full red line is a fit considering a constant yield . d) Switching yield versus sample bias (black dots). A typical $dI/dV$ spectrum of an open isomer (red curve) is added to the graph.}
\label{fig3}
\end{figure*}

Next, we aim at resolving the switching mechanism. We first determine the rate of the switching as a function of tunneling current. Because the switching is non-reversible, the rate is determined as shown exemplary in \Figref{fig3}a. While the current is kept constant, the $dI/dV$ signal and the tip height on the addressed molecule is monitored. A switching event is indicated by a sudden change of the $dI/dV$ signal and the tip height. To confirm the ring-closing reaction, we recorded $dI/dV$ spectra afterward. Only when the typical resonance structure of the closed isomer is observed, we take this event into account for the statistical analysis. The rate is then calculated from the time laps between parking the STM tip over the specific molecule and having set the specific bias voltage and the sudden change in tip-height and $dI/dV$ signal. We carry out this process for more than 200 molecules at different currents and different bias voltages.

In \Figref{fig3}b we plot the rate as a function of tunneling current for a voltage of 2~V. The current dependence of the switching rate follows a law in $I^{N}$, with $N=0.74 \pm 0.04$. A power $N=1$ would suggest a one-electron process, $N=2$ a two-electron process, etc. \cite{Kim2002, Lotze2012, Stipe1998, Iancu2006,Eigler1991}. An exponent less than one thus hints at a one-electron process mixed with an electron-independent process. This will be discussed in more detail in section \ref{sec_Efield}.
By converting the rate at a specific current into the electron yield for the switching, we obtain the plot in \Figref{fig3}c for a bias voltage of 2~V. The horizontal line labels a constant yield with current. We notice a small tendency of a decrease of yield with tunneling current, albeit a constant line lies almost within the error bars. We will discuss the origin of a possible deviation below.

We further evaluate the switching yield as a function of bias voltage in \Figref{fig3}d. Below 1~V, the yield is negligibly small. It increases to $2\times10^{-13}$ per electron at 1.2~V. A further rapid increase in the yield is observed at $\sim$~1.8~V and seems to level off again with a yield of $1\times10^{-9}$ per electron at 2.2~V. The shape of this drastic increase by four orders of magnitude follows the resonance structure as detected in $dI/dV$ spectra for the open isomer (red curve). Indeed, the threshold at $\sim$~1~V corresponds to the first resonance  and the more pronounced increase at $\sim$1.8~V is observed at the energy close to the second resonance. This behavior shows that the switching process is resonantly enhanced by electrons tunneling into the unoccupied molecular orbitals \cite{Henningsen2007, Henzl2010, Lotze2012, Lastapis2005}. We note also that in the negative bias regime, where we did not detect any molecular states of the open isomer, we also observe no switching.

\subsection{Electric field-induced switching} 
\label{sec_Efield}

\begin{figure}[ht]
\begin{center}
\includegraphics[width=1\columnwidth]{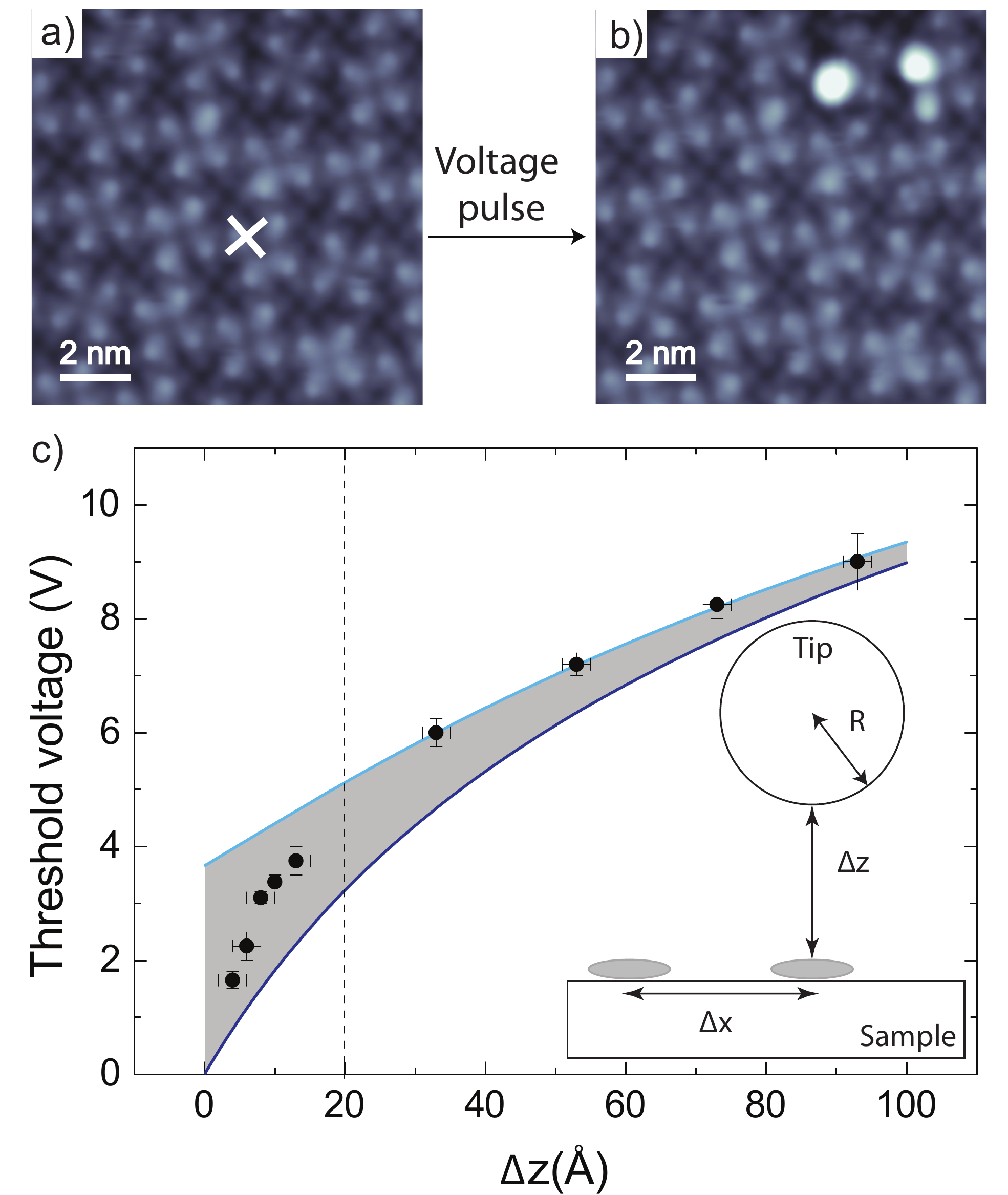}
\end{center}
\caption{ STM topographies before a) and after b) a voltage pulse has been applied at the position marked by the white cross with 9~V during 60~s and with the tip retracted by 9~nm from the tunneling condition (11~$\times$~11~nm$^{2}$, V~=~-1.3~V , I~=~50~pA). c) Evolution of the threshold voltage to observe switching events as shown in a)-b) versus $\Delta z$,  the tip-surface distance (black dot). The light blue line is a fitting for the points above 20~\AA\ with an electric field model, with the tip defined as a sphere of radius $R$~=~4.1~nm  and a distance $\Delta x$~=~5~nm as defined by the scheme in inset. The dark blue line is similar to the light blue one but with $\Delta x$~=~0~nm.}
\label{fig4}
\end{figure}

The observations in the previous section suggest an electron-induced process. However, the deviation of the switching rate from a strict proportionality with the current ($N < 1$) and the non-constant yield with current indicate that another parameter affects the switching process.

To determine this additional effect, we aim at separating it from the current-induced switching. For this, we carry out experiments by retracting the tip from the surface. In \Figref{fig4}a, the tip was withdrawn by 9~nm from the usual tunneling conditions, and a 9\,V-voltage pulse was applied during 60~s at the position marked by the white cross in the first topography. At this far distance, one can safely neglect the effect of any tunneling current. Nonetheless, some molecules are modified after the voltage pulse (\Figref{fig4}b). Note that these modifications are not only localized directly underneath the tip position during the pulse, but can appear in a radius of up to 5~nm. Subsequent $dI/dV$ spectra recorded on these molecules prove them to be switched into the closed isomer of the C5F-4Py. 

A non-local switching in the absence of tunneling current can be associated with the electric field present in the junction \cite{Alemani2006}. To confirm that, we determine the minimum voltage necessary to observe the switching as a function of the tip-sample distance, for a bias pulse of 60~s (black dots in \Figref{fig4}c).
 Usually, a simple plate-capacitor model is used to determine the electric field in the STM junction with $E=V/\Delta z$. This would yield a linear dependence of the threshold voltage with the tip sample distance, which is clearly not the case here. One may argue that a finite tunneling current may be responsible for the switching at small retraction distances. We therefore distinguish two regimes. The gray dashed line in \Figref{fig4}c at 20~\AA\ tip retraction separates the regimes of finite current and negligible current, i.e., a current below the fA range. If one extrapolates the threshold voltage by a linear dependence, the line does not cross zero voltage at tip-molecule contact. Hence, the plate-capacitor model is insufficient to explain the distance-dependence of the switching voltage.

A better model of the tip-sample junction may be a description by a spherical tip in front of a plate. The electric field in this geometry is not easy to calculate, but the $z$ component of the electrostatic field in the vicinity of the sphere may be approximated by \cite{Girard1993,Devel1995}:

\begin{equation}
 E= \frac{V}{\Delta z} \frac{\sqrt{D(1+D)}}{log(\sqrt{D}+\sqrt{1+D})} ,
\label{eq1}	
\end{equation}

with $D=\Delta z / R$, where $R$ is the radius of the tip, $V$ the bias voltage and $\Delta z$ is the tip-sample distance. From this formula we can therefore determine the required bias voltage to keep the electric field constant at a specific tip-sample distance. 

Note that we observed the switched isomers typically within a radius of 5~nm away of the position of the pulse ($\Delta x$ in the scheme of \Figref{fig4}c). This has to be considered in our model replacing in equation (\ref{eq1}) $\Delta z$ by $\sqrt{(R+\Delta z)^{2} +\Delta x ^{2}} - R $. 

Using this model, we fit our data for $\Delta z$ above 20 \AA\ and obtain the curve in light blue in \Figref{fig4}c with $\Delta x$ being fixed at 5~nm, while $E$ and $R$ were fitting parameters with final values of 0.21~V/\AA~ and 4.1~nm, respectively. Finally the dark blue line is obtained with the same $E$ and $R$ values but with $\Delta x=0$. All our data points fall into the range between the two curves. Hence, a switching mechanism involving the electric field is consistent with our observations, particularly for the events with a large tip-sample distance. 
For the case below 20~\AA, we see that the experimental points are closer to the fitting curve with $\Delta x=0$. This is in agreement with our observation that switching events with small $\Delta z$ are generally localized underneath the tip (see section \ref{sec_elinduced}). However, we also note that this regime also includes electron-induced switching events. 

Taking up on our observation that the rate scaled as $I^N$ with $N=0.74 \pm 0.04$, we can now also understand this sublinear behavior. An increase in tunneling current is accompanied by a smaller tip-molecule distance (at the same bias voltage). Hence, the electric field raises and more and more switching events due to the deformation of the potential energy landscape become possible. The crossing of the remaining barrier without electrons may be due to tunneling or the finite temperature. This explains that in average less than one electron is needed for triggering a switching process. This argument is directly connected and in line with a small decrease of the switching yield with increasing current. 

\subsection{Switching products}

\begin{figure}[ht]
\begin{center}
\includegraphics[width=1\columnwidth]{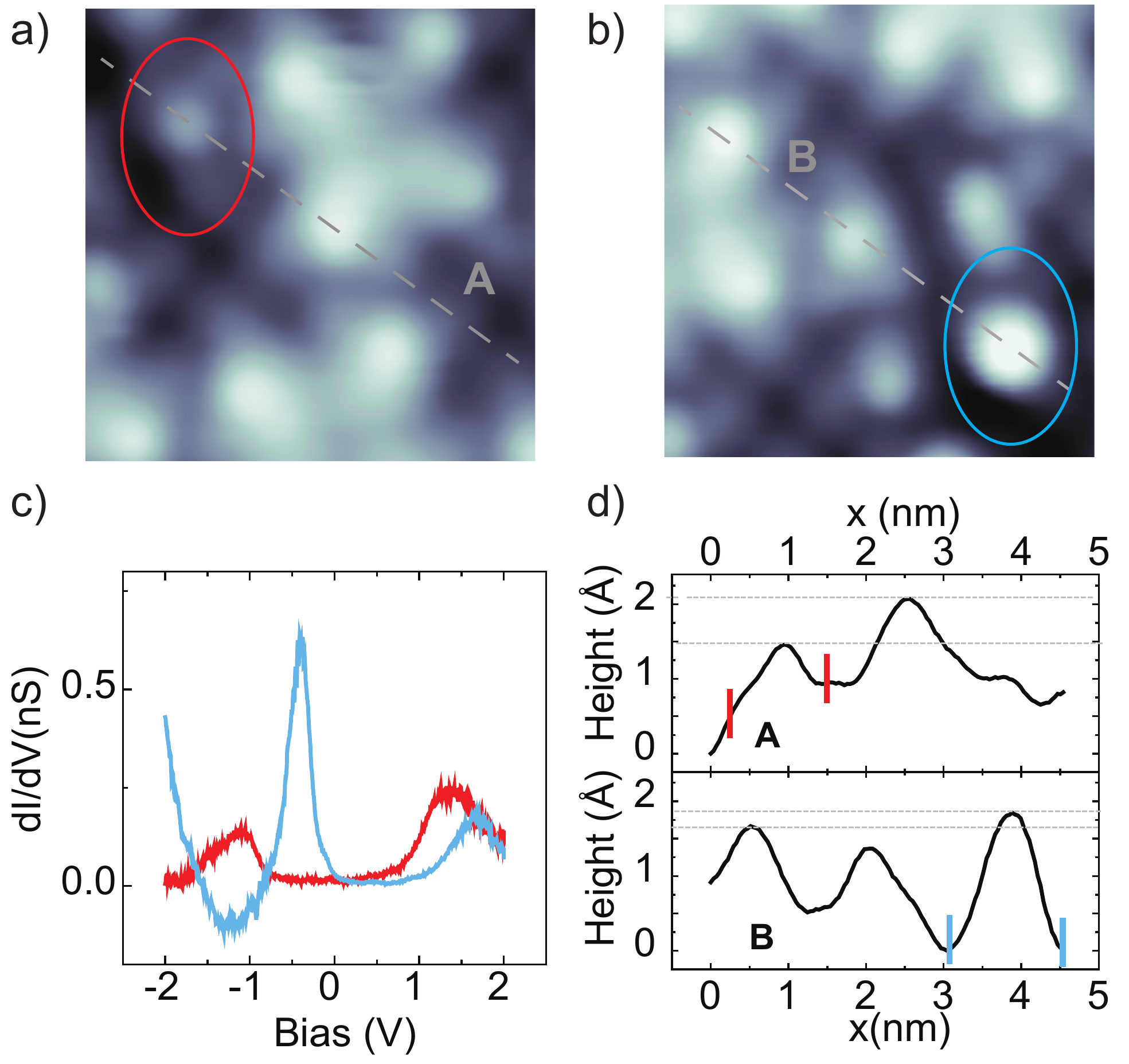}
\end{center}
\caption{ STM topographies with a molecule switched by electrons (a) and by the electric field (b) (4~$\times$~4~nm$^{2}$, V~=~0.2~V, I~=~50~pA). The switched molecules are framed by ellipses. c) $dI/dV$ spectra recorded on the top of the framed molecule in a) and b). d) Height profiles corresponding to the dashed line in a) (A) and in b) (B),respectively.} 
\label{fig5}
\end{figure}

In the previous sections, we discussed two different mechanisms, one with electrons and one with electric field, as a trigger for the ring-closing reaction in C5-4Py.  \Figref{fig5}a and b show two STM topographies with a molecule switched by electrons and in the electric field, respectively. In each image the corresponding molecule is framed by an ellipse. 
\Figref{fig5}c presents the $dI/dV$ spectra corresponding to these two molecules. The molecule switched with an electron (with 2~V and 200~pA), exhibits a spectrum (red curve) similar to the one shown in \Figref{fig2}c. The HOMO is found at -1.2~V and the LUMO at 1.2~V. The gap thus amounts to 2.4~V. For the molecule switched in the electric field with a pulse of 9.5~V and $\Delta z$~=~8~nm, the spectrum (blue curve) is distinctly different. The HOMO-LUMO gap is essentially the same (2.2~V) but the complete spectrum is shifted to larger positive bias voltages by $\sim$~0.6~V. Moreover, the width of the resonances are narrower (300~mV compared to 600~mV). Both differences can be explained by a change of the interaction between the molecule and the substrate. The closed isomer, which has been produced by the electric field exhibits a smaller electronic coupling. 
Additionally, the STM topographies in \Figref{fig5}a-b (recorded at 0.2~V, $i.e$ in the HOMO-LUMO gap to eliminate all electronic effects) reveal different apparent  heights of the molecules. While the electron-switched molecule is lower than the neighboring open molecules (height profile A in \Figref{fig5}d), the electric-field switched molecule is higher (height profile B). This observation supports our previous conclusion that with the electric field the closed isomer is more decoupled from the substrate.  One may speculate that the electric field lifts the molecule due to the presence of a dipole.

\section{Conclusions}
In summary, we have investigated the switching behaviour of C5F-4Py adsorbed on a Ag(111) surface by an STM tip. We could induce switching from the open to the closed form throughout a wide range of current and voltage parameters. We could identify an electron-induced  and an electric field-induced ring-closing reaction. The electron-induced switching yield is strongly enhanced by tunneling through unoccupied molecular orbitals. Below-resonance tunneling leads to a yield in the order of $10^{-14}$ per electron, while on-resonance tunneling pushes the yield up to $10^{-9}$ per electron. However, even in the absence of electron tunneling, switching can be observed. In this regime, the electric field modifies the barrier and leads to the ring-closing reaction. Our data also shows the limits of the commonly employed plate-capacitor model. The switching voltage can be reproduced by a model considering a spherical tip. While the analysis of switching rate, yield and distance dependence allowed us to draw a conclusive picture on the processes at work, we also noticed a difference in the adsorption state of the closed isomer depending on how the reaction proceeded. The molecules produced by the electric-field effect interact less with the substrate, presumably because they are slightly lifted from the surface. Our results highlight the interplay of electron- and field-induced switching, which is important in electronic devices.

\section{Acknowledgments}
We gratefully acknowledge funding by the German research foundation within the framework of the SFB 658, and the European Research Council for the ERC grant "NanoSpin". We thank Elke Scheer for suggesting the study of these molecules.

\bibliography{Switch_arxiv}

\end{document}